\documentclass[useAMS,usenatbib]{mn2e}
\usepackage{graphicx}

\title[A synchrotron superbublle in  IC\,10]
{A synchrotron superbubble in the IC 10 Galaxy: a hypernova
remnant?}
\author[T. A. Lozinskaya and A. V. Moiseev]{T. A. Lozinskaya$^{1}$\thanks{E-mail:
lozinsk@sai.msu.ru} and A. V.
Moiseev$^{2}$\\
$^{1}$Sternberg Astronomical Institute, Universitetskii pr. 13, Moscow, 119991 Russia\\
$^{2}$Special Astrophysical Observatory, Nizhnii Arkhyz,
Karachaevo-Cherkesia, 369167 Russia }

\begin{document}

\date{Accepted 2007 June 29. Received 2007  June; in original form 2007 June 4}

\pagerange{\pageref{firstpage}--\pageref{lastpage}} \pubyear{2007}

\maketitle

\label{firstpage}

\begin{abstract}
The nature of the synchrotron superbubble in the IC\,10 galaxy is
discussed using  the results of our investigation  of its ionized
gas structure, kinematics, and emission spectrum from observations
made with the 6-m telescope of the Special Astrophysical
Observatory of the Russian Academy of Sciences,  and based on our
analysis of the radio emission of the region. The hypernova
explosion is shown to be a more plausible mechanism of the
formation of the synchrotron superbubble compared with the earlier
proposed model of multiple supernova explosions. A compact remnant
of this hypernova may be identified with the well known X-ray
binary X-1 -- an accreting black hole.
\end{abstract}


\section{INTRODUCTION.}

The synchrotron superbubble in the IC\,10 galaxy was discovered by
\citet{y1}. They associated it with the explosion of about ten
supernovae. The synchrotron nature of the radio emission of this
superbubble is corroborated by the high degree of its polarization
\citep{c2}. The multiple supernova explosions model was also
adopted by \citet{b3} and \citet{t1}.

We provided a detailed study  of the structure, kinematics, and
emission spectrum of the ionized gas in the region of the
synchrotron superbubble based on  observations made with the 6-m
telescope of the Special Astrophysical Observatory of the Russian
Academy of Sciences (SAO RAS). We suggest, in accordance with our
observations and based on our analysis of the radio emission of
the region, that the synchrotron superbubble was produced by a
hypernova explosion and not via multiple supernova explosions as
was believed until now.

\section{RESULTS OF OBSERVATIONS.}

We observed the ionized gas in the synchrotron superbubble region
with the SAO RAS 6-m telescope using SCORPIO focal reducer
\citep{afan05} operating in three modes: direct [SII]-lines
images, long-slit spectroscopy, and observations with a scanning
Fabry--Perot interferometer (FPI) in the H$\alpha$  line. We
report the detailed results of our observations in a separate
paper \citep{l1}. In this Letter we summarize the main results of
these observations and the ensuing conclusions.

\begin{figure}
\includegraphics[width=8.5cm]{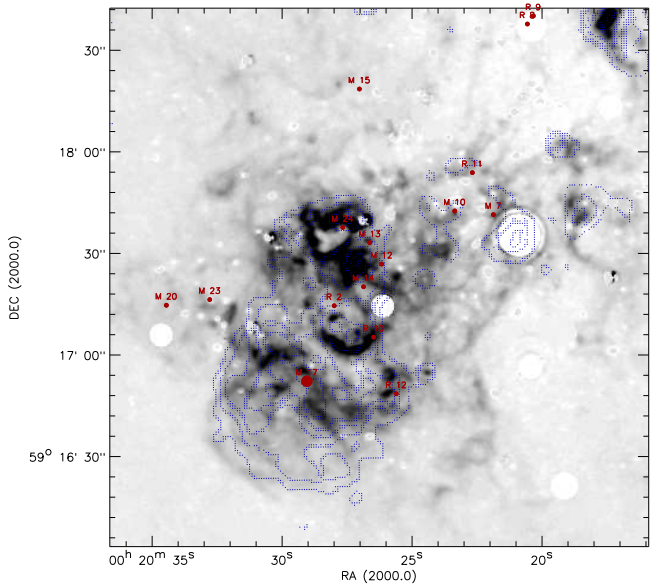}
\caption{The [SII]6717+31\AA\, lines image of the region taken with the
6-meter telescope of the SAO RAS with SCORPIO focal reducer. Small red
circles show spectroscopically confirmed WR stars. The large red circle in
the South-eastern part of the region is the WR star M17, a component of the
brightest compact X-ray source X-I in the IC\,10 galaxy (see text). The blue
lines show the superimposed 20-cm continuum radio emission contours from
\citet{y1}.} \label{fig1}
\end{figure}

Fig.~\ref{fig1} shows the resulting [SII]6717+31\AA\, lines image of the
region with 20-cm continuum radio emission contours from \citet{y1}
superimposed. Compared to H$\alpha$ images (from \citet*{gil03} or  from our
FPI data), our continuum-subtracted [SII] image reveals the most
well-defined filamentary shell  which one can identify with the synchrotron
superbubble.

The size of the filamentary [SII] shell is about $44''$, which
corresponds to 170 pc at the distance of 790 kpc \citep*{v1}. Its
central coordinates $[\alpha_{(2000)} = 0^{h}20^{m}29^{s}$,
$\delta_{(2000)}=59^{\circ}16'40'']$ agree with those of the radio
shell observed by \citet{y1}.

\begin{figure}
\includegraphics[width=8.5cm]{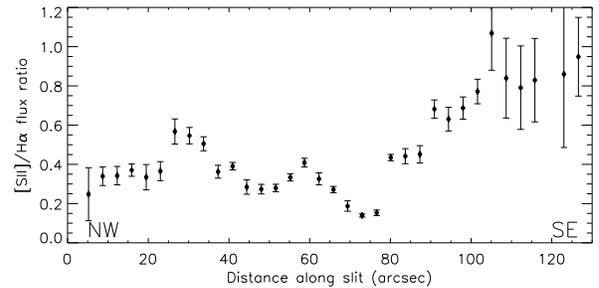}
\caption{The variations of [SII]/H$\alpha$ flux ratio along the
spectrograp slit. The  position angle of the slit was
$PA=133^\circ$. The synchrotron superbubble correspods to the
distances $90-130$ arcsecs.} \label{fig2}
\end{figure}

Our long-slit spectra are  indicative of the enhanced [SII]
emission in the synchrotron superbubble area, much stronger than
in other star-forming regions in the galaxy.  Indeed, the
I([SII])/I(H$\alpha$) ratio in the superbubble lies in the
$0.6-1.0$ interval (see Fig.~\ref{fig2}) and this is consistent
with the corresponding ratios of supernova remnants (SNRs). Fig.~1
in \citet{r1} and fig.~4 in \citet{h1} also point to the bright
[SII] emission in the region.

We have performed a very detailed H$\alpha$ line study of the
kinematics of ionized gas using  a  scanning FPI, and analysed
more than 40 position-velocities (P-V) diagrams crossing the
synchrotron superbubble in various directions. The  FPI data
allowed us to estimate the characteristic expansion velocity of
the system of bright knots and filaments to be 50 -- 80
km\,s$^{-1}$. The measured expansion velocity fully agrees with
the 50--70 km\,s$^{-1}$ value mentioned by \citet{b3}.

We   use the above expansion velocity, combined with the electron density of
$n_{e} \simeq 20-30$ cm$^{-3}$ estimated from the [SII]6717/6731\AA\,
emission lines ratio, to evaluate the mass and kinetic energy of the optical
shell to be about $M\simeq 8\times 10^{38}$ g and $E_{kin}\simeq (1-3)\times
10^{52}$ erg, respectively.   (As is commonly adopted for SNRs, we assume
that the shell thickness is about 0.1 of its radius.)

The energy obtained is between the  value of $E_{kin}\simeq (5-6)\times
10^{52}$ erg estimated by \citet{t1} from the mean halfwidth of the
H$\alpha$ line in the synchrotron superbubble, and $E_{kin}\simeq
(0.6-1.2)\times 10^{51}$ erg reported by \citet{b3} and \citet{r2}.

\begin{figure}
\includegraphics[width=8.5cm]{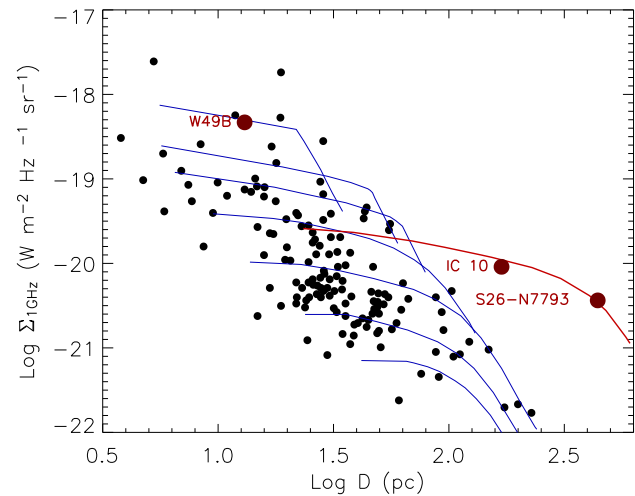}
\caption{The $\Sigma_{1GHz}-D$ diagram  mainly based on  data extracted from
fig.~6 in \citet*{a1}. The black dots show the position of  SNRs  in our
Galaxy and several nearby galaxies. The red circles  mark the position of
the hypernova remnant  candidates including the synchrotron superbubble in
IC\,10. The red line is the theoretical $\Sigma-D$ relation constructed by
\citet*{a1} for the hypernova explosion with the initial energy of E$_{)} =
5\times 10^{52}$ erg in a medium with unperturbed density $n_{0}=0.01$
cm$^{-3}$. The blue curves in the figure show the $\Sigma-D$ dependences for
supernova remnants with standard energy E$_{0} = 10^{51}$ erg expanding in
media of different densities.} \label{fig3}
\end{figure}

\section{THE NATURE OF THE SUPERBUBBLE}

The kinetic energy of an old SNR is lower than about 30 per cent of the SN
explosion energy \citep{c1}. Thus our inferred kinetic energy for the
optical shell in the synchrotron superbubble corresponds to the explosion of
about dozen supernovae plus stellar winds of their host OB association, or to
a hypernova explosion.

However, our analysis of the synchrotron radiation of the superbubble leads
us to suggest that a hypernova explosion explains better the nature of this
radiation than do multiple supernovae.

First, the surface  brightness $\Sigma_{(1GHz)} = 10^{-20}$ W m$^{-2}$
Hz$^{-1}$ sr$^{-1}$  for  the size of superbubble of  170 -- 190 pc  fits
perfectly well the theoretical $\Sigma(D)$--dependence that \citet{a1}
derived for a hypernova explosion with the initial energy of E$_{0} =
5\times 10^{52}$ erg in a medium with density $n_{0}=0.01$ cm$^{-3}$ (see
Fig.~\ref{fig3} where we show the superbubble in IC\,10  by a red circle).

Second, and this is most important, we are the first to allow for the fact
that explosions of `recent' supernovae in the model of \citet{y1} occur in a
tenuous cavity inside the common shell swept out by the `first' supernovae.

The first few SN explosions may indeed have produced several times
stronger radio emission compared to that of a single SN. However,
the situation changes radically as the first SNe create the
common swept-out supershell. The remnants of subsequent SNe
expand in a low-density medium inside the swept-out cavity, and
their radio emission rapidly decays because of the adiabatic
expansion of the cloud of relativistic particles with magnetic
field.

The radio brightness of a SNR depends on the parameters of the ambient medium
as
 $$\Sigma(D) \propto n_{0}^{2/3} B_{0}^{1.5} \propto n_{0}^{2/3 +1.5 k_{0}},$$
where $B_{0} \propto n_{0}^{k_{0}}$ \citep{a1}.

The brightness of an SNR in the cavity where the density is 10 or 100 times
lower than the ambient density can be easily seen to decrease by a factor of
5 or 20, respectively.

The allowance for the fact that the interstellar magnetic field is frozen in
the gas further strengthens this conclusion, because the magnetic field
inside the cavity is weaker than the ambient field. Correspondingly, the
surface radio brightness of the SNR in the tenuous cavity mentioned above
decreases by a factor of several tens or several hundred.

Of course, these are just qualitative estimates, because some SNe may explode
in the dense medium near the swept-out shell.

We nevertheless conclude that  subsequent SN explosions in the
model of \citet{y1} contribute little to the radio brightness of
the synchrotron shell created by the first SN explosions, implying
a further increase of the required number of supernovae. That is
why we believe a hypernova explosion to be a more plausible
mechanism for the formation of the synchrotron superbubble than
multiple supernova explosions.

The age of this hypernova remnant determined by its size and the expansion
velocity of 50-80 km\,s$^{-1}$ inferred in this paper corresponds to $t
\simeq (4-7)\times 10^{5}$ yr if it is at the Sedov stage or $t \simeq
(3-5)\times 10^{5}$ yr for the case of the radiative cooling stage.

Such an age also supportive of the hypernova  hypothesis, because ten
supernova explosions require about  2 orders of magnitude longer time.

The anomalously high number of  Wolf-Rayet (WR) stars in the galaxy IC\,10
provides indirect evidence in support of the same conclusion. In the case of
a normal IMF the anomalously high density of WR stars implies virtually
`simultaneous' current burst of star formation covering most of the galaxy
(see, e.g., \citet{m1} and references therein). We `caught'  IC\,10 during
the short stage of its evolution when massive WR progenitors are still
alive, but the supermassive pre-hypernova has already finished its life, and
we observe the remnant of its explosion as the  synchrotron superbubble.

One can believe  the compact remnant of this hypothetical hypernova to
coincide with the brightest X-ray source in the galaxy X-I, discovered by
\citet{b2}. X-I  is a stellar-mass black hole accreting from WR star M17; the
mass of this black hole is $\simeq 4 M_\odot$ if it is not spinning, or up to
$\approx 6$ times higher if there is significant spinning \citep{b1,w1}.

If the synchrotron shell is indeed a hypernova remnant, one would expect to
find an extended X-ray emission in the region. Based on \textit{CHANDRA}
observations, \citet{b1}  found evidence for faint extended X-ray emission
`cospatial' with the synchrotron superbubble. However, accurate reduction of
\textit{XMM-Newton} and \textit{CHANDRA} observations (removing the X-ray
CCD-readout streaks of X-I) led \citet{w1} to conclude that  the faint
diffuse thermal X-ray emission  appears to be associated with the intense
star-forming region. New X-ray observations are highly desirable.

\section{DISCUSSION AND CONCLUSION.}

Compared to the other two hypothetical hypernova remnants --- W49B and
S26-N7793 --- shown in Fig.~\ref{fig3} the synchrotron supershell in IC 10
appears to be the most confidently identified object. Indeed as \citet{k1}
show in their paper, the progenitor of W49B was a supermassive star. At the
same time, the location of the radio source in the $\Sigma-D$ relation
agrees excellently with the results of the computations that \citet{a1}
performed for a SN remnant with a standard energy of $E_{0} = 10^{51}$ erg
in a medium with density $n_{0} = 5$ cm$^{-3}$.

The  hypernova  remnant S26-N7793 in the NGC7793 galaxy was also
earlier attributed to multiple supernova explosions (see
\citet{p1} and references therein). This S26-N7793 remnant is most
probably at a later evolutionary stage than the Synchrotron Shell
in IC\,10. However, we still do not understand its shape: it
appears as an SNR with a long filament as an extension.

NGC 5471B in the galaxy M\,101 is  one of most reliably identified hypernova
remnant \citep{wang} and was studied in  detailed in radio, optical and X-ray
ranges (see \citet{s1}, \citet{chu}, \citet{chen} and also references
therein). Its  kinetic energy reaches  $E_{kin} =5\times 10^{51}$ erg
($E_{0} \ge 10^{52}$ erg), kinematic age is no more than $10^5$ yr, it is
characterized by high [SII]/H$\alpha$ ratio. The one problem  of its
identification is that NGC 5471B lies in an active star formation region  in
the giant  HII complex NGC 5471 and contains a large number of faint
clusters and two clusters as rich as R136 within the bright [SII]-shell  NGC
5471B \citep{chen05}.

The synchrotron superbubble in the IC\,10 does not have this difficulty.
\citet{hunter} has distinguished two clusters near the southern border of
the superbuble: 4-6 and 4-7. However these are not richest clusters in the
galaxy and they could not host dozen supernova explosions.

Note in conclusion that the  synchrotron superhshell in IC\,10 can be
identified as a hypernova remnant based on a combination of several criteria
--- first, very high kinetic energy of the shell; second, the presence of a
bright extended spherically symmetric source of synchrotron radio emission,
which is difficult to explain by multiple supernova explosions; third, the
optical shell with high [SII]-line brightness, which is ``cospatial'' with
the radio source and has a kinematical age of $t \simeq (3-7)\times 10^{5}$
yr, and, fourth, the  presence of a compact remnant of the explosion of a
very massive star.

\section*{ACKNOWLEDGMENTS}

We thank Alexander Burenkov and Azamat Valeev for their help in observations
and also Gennadij Bisnovatyj-Kogan, Konstantin Postnov, and Sergej Popov for
fruitful discussions. We are grateful to the anonymous referee for useful
comments. This work was supported by the Russian Foundation for Basic Research
(projects nos.~ 05--02--16454, and 07--02--00227) and is based on the
observational data obtained with the 6-m telescope of the Special
Astrophysical Observatory of the Russian Academy of Sciences funded by the
Ministry of Science of the Russian Federation (registration number 01-43).


\begin{thebibliography}{99}

\bibitem[\protect\citeauthoryear{Afanasiev \& Moiseev}{2005}]{afan05}
Afanasiev V.L., Moiseev A.V., 2005, Astronomy Letters, 31, 194
(astro-ph/0502095)

\bibitem[\protect\citeauthoryear{Asvarov}{2006}]{a1}
Asvarov A.I., 2006, A\&A, 459, 519

\bibitem[\protect\citeauthoryear{Bauer \& Brandt}{2004}]{b1}
Bauer F.E., Brandt W.N., 2004,  ApJ, 601, L67

\bibitem[\protect\citeauthoryear{Brandt et al.}{1997}]{b2} Brandt
W.D., Ward M.J., Fabian A.C., Hodge P.W., 1997, MNRAS, 291, 709

\bibitem[\protect\citeauthoryear{Bullejos \& Rozado}{2002}]{b3} Bullejos A.,
Rozado M., 2002, Rev.Mex.A.A.  (Serie de Conferencias), 12,  254

\bibitem[\protect\citeauthoryear{Chen et al.}{2002}]{chen}
Chen C.-H.R., Chu  Y.-H., Gruendl R., Lai S.P., Wang Q.D., 2002, AJ, 123, 2462

\bibitem[\protect\citeauthoryear{Chen et al.}{2005}]{chen05}
Chen  C.-H.R., Chu Y.-H., Johnson K.E., 2005, ApJ, 619, 779

\bibitem[\protect\citeauthoryear{Chevalier}{1974}]{c1}
Chevalier R.A., 1974, ApJ, 188, 501

\bibitem[\protect\citeauthoryear{Chu \& Kennicutt}{1986}]{chu}
Chu Y.-H., Kennicutt R.C., 1986, ApJ, 311, 85

\bibitem[\protect\citeauthoryear{Chyzy et al.}{2003}]{c2} Chyzy K.T.,
Knapik J., Bomans D.J., Klein U., Beck R., Soida M., Urbanik M., 2003, A\&A,
405, 513

\bibitem[\protect\citeauthoryear{Gil de Paz,  Madore \& Pevunova}{Gil de Paz
et al.}{2003}]{gil03} Gil de Paz A.,  Madore B.F., Pevunova O., 2003, ApJS,
147, 29

\bibitem[\protect\citeauthoryear{Hidalgo-Gamez}{2005}]{h1}
Hidalgo-Gamez A.M., 2005, A\&A, 442, 443

\bibitem[\protect\citeauthoryear{Hunter}{2001}]{hunter}
Hunter D.A., 2001, ApJ,  559, 225

\bibitem[\protect\citeauthoryear{Keohane et al.}{2006}]{k1}
Keohane J.W., Reach W.T., Rho J., Jarrett T.H., 2006, ApJ, 654, 938

\bibitem[\protect\citeauthoryear{Lozinskaya et al.}{2007}]{l1} Lozinskaya  T.A.,
Moiseev  A.V., Podorvanyuk  N.Yu., Burenkov  A.N., 2007, submitted
to Astronomy Letters

\bibitem[\protect\citeauthoryear{Massey et al.}{2007}]{m1} Massey  P.,
Olsen K., Hodge P., Jacoby G.H., McNeill R., Smith R., Strong  Sh., 2007, AJ,
133, 2393

\bibitem[\protect\citeauthoryear{Pannutti et al.}{2002}]{p1}
Pannuti T.G., Duric N., Lacey C.K., Ferguson A.M.N., Magnor M.A., Mendelowitz
C., 2002, ApJ, 565, 966

\bibitem[\protect\citeauthoryear{Rosado et al.}{1999}]{r1}
Rosado M.,  Bullejos A., Valdez M., Georgiev L., Lacey C., Borissova J.,
Esteban C., 1999, in Y.-H. Chu, N. Suntzeff, J. Hesser,  D. Bohlender, eds.,
Proc. IAU Symp. 190,  New Views of the Magellanic Clouds, Astron. Soc. Pac.,
San Francisco, p. 168

\bibitem[\protect\citeauthoryear{Rosado et al.}{2002}]{r2}
Rosado M., Valdez-Gutierrez M., Bullejos A., Arias L., Georgiev L.,
Ambrocio-Cruz, P., Borissova J., Kurtev R., 2002, in Margarita Rosado, Luc
Binette, and Lorena Arias, eds., ASP Conf. Ser. 282,  Galaxies: The Third
Dimension, Astron. Soc.  Pac., San Francisco, p. 50

\bibitem[\protect\citeauthoryear{Skillman}{1985}]{s1}
Skillman E.D., 1985, ApJ, 290, 449

\bibitem[\protect\citeauthoryear{Thurow \& Wilcots}{2005}]{t1}
Thurow  J.C., Wilcots  E.M., 2005, AJ, 129, 745

\bibitem[\protect\citeauthoryear{Vacca, Sheehy \& Graham}{Vacca et al.}{2007}]{v1}
Vacca  W.D., Sheehy  C.D., Graham  J.R., 2007, ApJ,  662, 272

\bibitem[\protect\citeauthoryear{Wang}{1999}]{wang} Wang  Q.D., 1999, ApJ, 517, L27

\bibitem[\protect\citeauthoryear{Wang, Whitaker \& Williams}{Wang  et al.}{2005}]{w1}
Wang  Q.D., Whitaker  K.E., Williams  R., 2005, MNRAS, 362, 1065

\bibitem[\protect\citeauthoryear{Yang \& Skillman}{1993}]{y1}
Yang  H., Skillman  E.D, 1993, AJ, 106, 1448

\end{thebibliography}
\end{document}